\documentclass[fleqn,usenatbib,useAMS]{mnras}

\usepackage{graphicx}	
\usepackage{amsmath}	
\usepackage{amssymb}	
\usepackage{multicol}        
\usepackage{booktabs}
\usepackage{float}
\usepackage[small,labelfont=bf,up,textfont=it,up]{caption}
\usepackage{bm}		
\usepackage{pdflscape}	

\usepackage{tabularx}
\usepackage{longtable}

\usepackage{xspace}
\usepackage{soul}

\makeatletter
\newcommand{\@todonotes@enable}{1}
\newcommand{\@todonotes@inline}{1}
\makeatother
\input{notes.tex}

\usepackage[T1]{fontenc}
\usepackage{ae,aecompl}

\usepackage{newtxtext,newtxmath}

\title[Tidal Tails of Globular Clusters]{The Effect of Dwarf Galaxies on the Tidal Tails of Globular Clusters}

\author[El-Falou \& Webb]{Nada El-Falou$^1$ \thanks{E-mail:  nada.elfalou@mail.utoronto.ca (NF), webb@astro.utoronto.ca (JW)} \& Jeremy J. Webb$^1$ \\
$^1$Department of Astronomy and Astrophysics, University of Toronto, 50 St. George Street, Toronto, ON, M5S 3H4, Canada \\
}

\date{}

\pubyear{2021}

\begin{document}
\label{firstpage}
\pagerange{\pageref{firstpage}--\pageref{lastpage}}
\maketitle

\begin{abstract}

The tidal tails of globular clusters have been shown to be sensitive to the external tidal field. We investigate how Galactic globular clusters with observed tails are affected by satellite dwarf galaxies by simulating tails in galaxy models with and without dwarf galaxies. The simulations indicate that tidal tails can be subdivided into into three categories based on how they are affected by dwarf galaxies: 1) dwarf galaxies perturb the progenitor cluster's orbit (NGC 4590, Pal 1, Pal 5), 2) dwarf galaxies perturb the progenitor cluster's orbit and individual tail stars (NGC 362, NGC 1851, NGC 4147, NGC 5466, NGC 7492, Pal 14, Pal 15), and 3) dwarf galaxies negligibly affect tidal tails (NGC 288, NGC 5139, NGC 5904, Eridanus). Perturbations to a cluster's orbit occur when dwarf galaxies pass within its orbit, altering the size and shape of the orbital and tail path. Direct interactions between one or more dwarf galaxies and tail stars lead to kinks and spurs, however we find that features are more difficult to observe in projection. We further find that the tails of Pal 5 are shorter in the galaxy model with dwarf galaxies as it is closer to apocentre, which results in the tails being compressed. Additional simulations reveal that differences between tidal tails in the two galaxy models are primarily due to the Large Magellanic Cloud. Understanding how dwarf galaxies affect tidal tails allows for tails to be used to map the distribution of matter in dwarf galaxies and the Milky Way.

\end{abstract}




\section{Introduction}

The tidal tails of a globular cluster are populated by stars that have escaped the cluster due to tidal stripping. A wide range of internal (e.g. two-body relaxation) and external (e.g tidal shocks) mechanisms result in individual stars gaining energy and migrating outwards towards the cluster's tidal radius (also referred to as the Jacobi radius) \citep{spitzer58,spitzer87,Heggie03,gieles06,kruijssen11}. Once a star gains enough energy to become energetically unbound from its host cluster, it will remain within the cluster's tidal boundary until it reaches either the cluster's L1 or L2 Lagrange points and escapes \citep{lee87, Ross1997, Fukushige2000}. Tidally stripped stars escape through their host cluster's Lagrange points with velocities that are either slightly greater or less than the cluster's velocity, such that their resulting orbit in the external tidal field of the galaxy are quite similar to the host cluster and other tidally stripped stars. Hence tidal tails are observed as kinematically cold streams of stars that trail and precede their progenitor globular cluster \citep{Grillmair16}.

The stellar density within tidal tails is low enough such that gravitational interactions between tail stars are minimal. Hence the properties of a tidal tail are determined by how stars escape the host cluster \citep[e.g.][] {Kupper2012, Webb2021} and the gravitational effect of the external tidal field. The relationship between the properties of a tidal tail and the host galaxy has been shown to be quite complex, with baryonic and non-baryonic substructure and the Milky Way's bar playing roles in altering the stellar density and velocity dispersion along the tail \citep{Ibata2002, Johnston2002, Amorisco16, pearson17, Bovy17, Banik2019}. The tail path itself, however, only depends on the macroscopic distribution of matter throughout the host galaxy. This dependence has made tidal tails, and streams in general, effective tracers of orbits in a given external field and allows them to be used as tools to help map the gravitational potential of the Milky Way \citep{bovy16_mwmap}.

Several recent studies have found that it is not only the gravitational field of a tail's host galaxy that is responsible for the shape of the tail path, but that satellite galaxies may also play a role as well. The  Magellanic Clouds have been shown to influence the orbital paths of other satellites of the Milky Way \citep{Patel20, Ji2021, Erkal2020LMC}, with the tails of the Sagittarius dwarf galaxy clearly being perturbed by the Large Magellanic Cloud \citep{Gomez_2015, Vasiliev2021}. Similarly, the paths of the Tucana and Orphan streams can only be reproduced if interactions with the Large Magellanic Cloud are taken into consideration \citep{Erkal2018, Erkal2019}. In addition to affecting the orbital path of streams, interactions with dwarf galaxies may also be the source of spurs as \citet{deBoer20} and \citet{li2021} find that Sagittarius may be responsible for the GD-1 spur \citep{price-whelan2018}.

With respect to globular clusters, \citet{Garrow2020} explored how the orbits of each Galactic cluster is affected by including the presence of six dwarf galaxies (Large Magellanic Cloud, Small Magellanic Cloud, Sagittarius, Draco, Ursa Minor, and Fornax) in the external tidal field.  The authors find that dwarf galaxies can cause the semi-major axis and eccentricity of a globular cluster to evolve significantly over time, with semi-major axes and eccentricities fluctuating on the order of $10\%$ and $4\%$ respectively over 12 Gyr. Outer clusters, with semi-major axes greater than 10 kpc, were more strongly affected by the presence of dwarf galaxies as they are more likely to have a dwarf pass within their orbit or even have a close interaction with a dwarf than inner clusters. When modelling the evolution of select globular clusters in external tidal fields, \citet{Garrow2020} further finds that the time evolution of a cluster's orbit due to the presence of dwarf galaxies can affect its mass loss rate relative to when it orbits in a Milky Way-like tidal field only. These results should also be reflected in the tidal tails of these clusters.

The purpose of this study is to identify which of the Galactic globular clusters with observed tidal tails will be most affected by dwarf galaxies. \citet{Piatti2020} lists 14 globular clusters from the literature that have observed tidal tails, which include NGC 288 \citep{Kaderali2019, Shipp2018}, NGC 362 \citep{Carballo2019}, NGC 1851 \citep{Shipp2018, Carballo2018}, NGC 4147 \citep{Jordi2010}, NGC 4590 \citep{Palau2019}, NGC 5139 \citep{Simpson2020, Ibata2019}, NGC 5466 \citep{Jordi2010, Belokurov2006}, NGC 5904 \citep{Grillmair2019}, NGC 7492 \citep{Navarrete2017}, Eridanus \citep{Myeong2017}, Pal 1 \citep{Niederste2010}, Pal 5 \citep{Odenkirchen2001, Starkman2020}, Pal 14 \citep{Sollima2011}, and Pal 15 \citep{Myeong2017}. Given the positions and velocities of each cluster \citep{Vasiliev19}, we generate model tidal tails using a particle spray method \citep{Fardal2015, Banik2019} in galaxy models with and without dwarf galaxies to determine how the tail path and the properties of the tidal tail itself are affected.

The method that we use to generate artificial tidal tails in different galaxy models is outlined in Section \ref{s_methods}. In Section \ref{s_results} we illustrate how satellite galaxy interactions alter the tidal tail path of each globular cluster. The effect of dwarf galaxy interactions on each globular cluster and tidal tail is then quantified in Section \ref{s_discussion}, where we discuss differences between tails in galaxy models with and without satellite galaxies. We summarize our findings in Section \ref{s_conclusion}.

\section{Methodology}\label{s_methods}

To observe and quantify the effects of neighbouring dwarf galaxies on the properties of a globular cluster's tidal tails, we make use of a particle spray method to simulate tidal tails in two galaxy models. The first galaxy model consists of only the Milky Way (MW galaxy model), while the second consists of the Milky Way and several dwarf galaxies (DG galaxy model). In this project, the Milky Way gravitational potential was assumed to equal the \texttt{MWPotential2014} model from \citet{galpy}.

The satellite dwarf galaxies included in the simulation are the Large Magellanic Cloud, the Small Megallanic Cloud, Draco, Ursa Minor, Fornax, Sculptor and Sagittarius. For each dwarf galaxy it is necessary to determine how its gravitational potential adds to the gravitational field of the Milky Way as a function of time. The masses and scale sizes of the Large Magellanic Cloud ($1 \times 10^{11} M_{\odot}$, $10.2$ kpc) \citep{Besla10, Laporte18}), Small Magellanic Cloud ($2.6 \times 10^{10} M_{\odot}$, $3.6$ kpc) \citep{Besla10}, Draco ($6.3 \times 10^{9} M_{\odot}$, $7.0$ kpc) \citep{Penarrubia08}, Ursa Minor ($2.5 \times 10^{9} M_{\odot}$, $5.4$ kpc) \citep{Penarrubia08}, Fornax ($2.0 \times 10^{9} M_{\odot}$, $3.4$ kpc) \citep{Goerdt06}, and Sagittarius ($1.4 \times 10^{10} M_{\odot}$, $7.0$ kpc) \citep{Laporte18} are used to set the gravitational potential of each dwarf galaxy equal to a Hernquist Potential \citep{Hernquist1990}. The Sculptor dwarf galaxy, on the other hand, was taken to be a Plummer Potential based on \citet{Battaglia2007} and \citet{ Battaglia2008}. The positions and velocities of each satellite are taken from mean values of \citep{Helmi} and their orbits were integrated backwards in the \texttt{MWPotential2014} model of the Milky Way for 5 Gyrs while accounting for dynamical friction using \texttt{galpy}\footnote{http://github.com/jobovy/galpy} \citep{galpy}. Hence we create a time-dependant gravitational potential that includes both the Milky Way and seven satellite dwarf galaxies. It is worth noting that the gravitational potentials of the dwarf galaxies were used on the globular clusters, but did not affect the Milky Way itself.

For a given globular cluster,the formation and evolution of its tidal tails was simulated using the particle spray method of \citet{Fardal2015}. In this method, tails are modeled as a collection of particles leaving the progenitor while the masses of the progenitor and stream are continually updated. We specifically make used of  \texttt{streamspraydf}, which is an implementation of the \citet{Fardal2015} mock tidal stream generator in \texttt{streamtools}\footnote{https://streamtools.readthedocs.io/en/latest/} that was first used in \citet{Banik2019}.  

In order to generate sufficiently long tidal tails, we assume each globular cluster has been losing mass for 5 Gyr. Hence the orbit of a given cluster is integrated backwards in both the MW and DG galaxy models, with the dwarf galaxies orbiting around the Milky Way in the latter case. The tails are then simulated as the cluster moves forward in time, eventually reaching its current place in the galaxy. This process was performed for globular clusters NGC 288, NGC 362, NGC 1851, NGC 4147, NGC 4590, NGC 5139, NGC 5466, NGC 5904, NGC 7492, Eridanus, Pal 1, Pal 5, Pal 14, and Pal 15. As discussed in the Introduction, these 14 globular clusters have all been observed to have tidal tails. The mass of each cluster, which is necessary to include when generating mock tidal tails, is taken from \citet{Baumgardt18}.

\section{Results}\label{s_results}

To quantitatively analyse the effect of dwarf galaxies on tidal tails, we compare the spatial distribution of each globular cluster's tail stars in both galaxy models. Comparisons are done in Galactocentric coordinates and the plane of the sky, allowing us to identify differences between tidal tails in the DG galaxy model and the MW galaxy model.

\begin{figure*}
    \includegraphics[width=\textwidth]{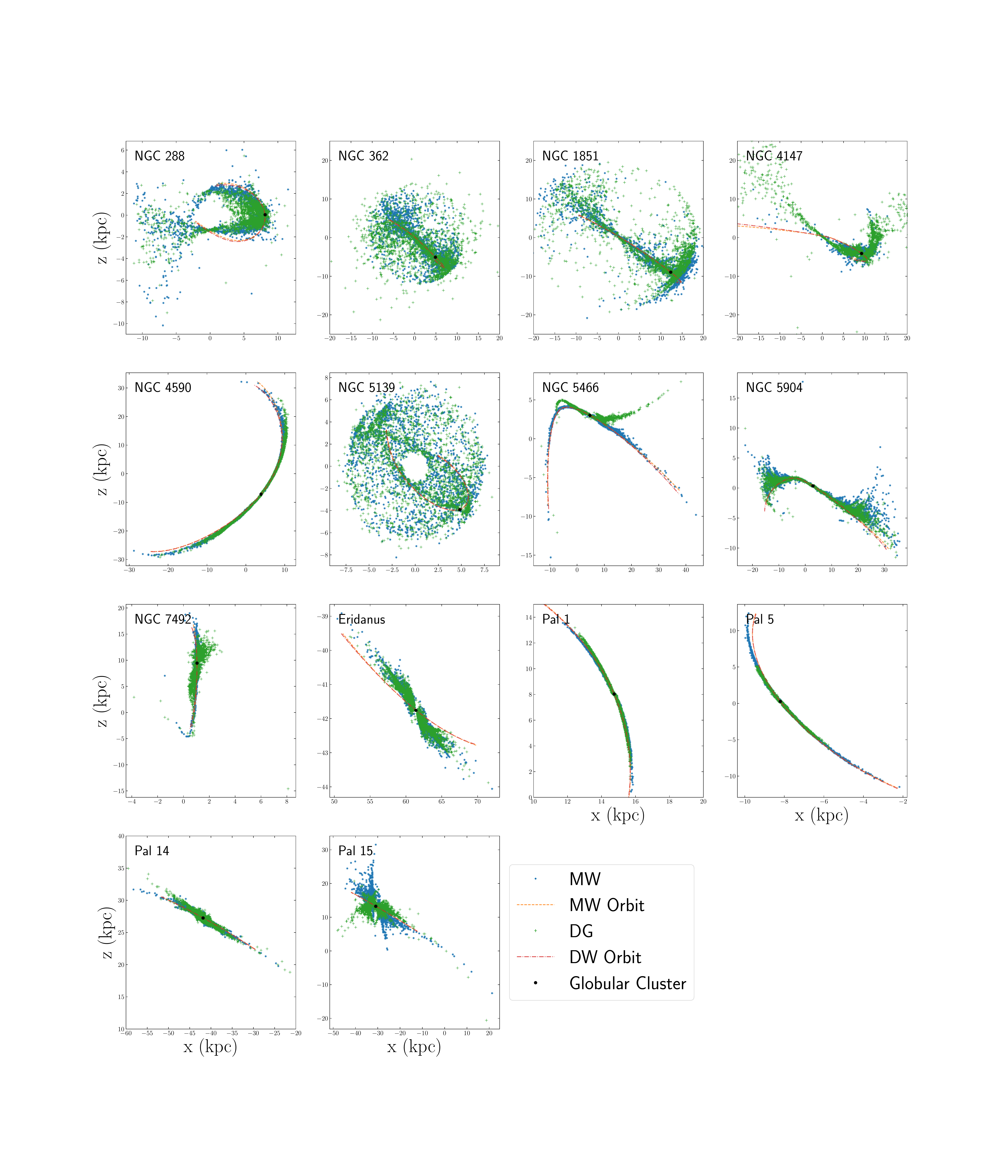}
    \caption{X-Y positions of all simulated tidal tail stars in the MW galaxy model (blue dots) and the DG galaxy model (green crosses) in Galactocentric coordinates. The location of the cluster is marked in black, while the orbital paths of each cluster is also illustrated for when the cluster orbits in the MW galaxy model (orange dashed lines) and the DG galaxy model (red dash dotted lines).}
    \label{fig:gc_xyall}
\end{figure*}

\begin{figure*}
    \includegraphics[width=\textwidth]{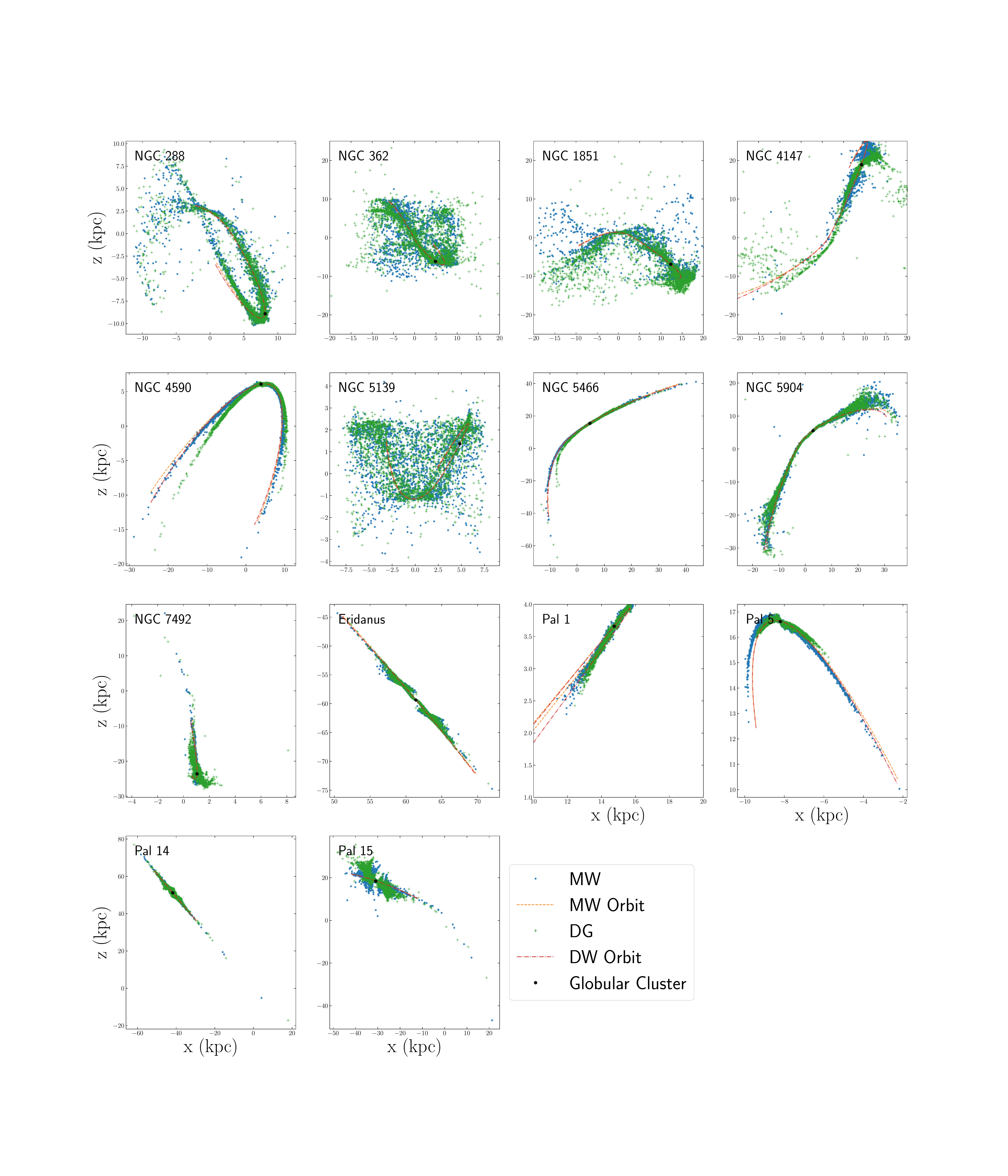}
    \caption{X-Z positions of all simulated tidal tail stars in the MW galaxy model (blue dots) and the DG galaxy model (green crosses) in Galactocentric coordinates. The location of the cluster is marked in black, while the orbital paths of each cluster is also illustrated for when the cluster orbits in the MW galaxy model (orange dashed lines) and the DG galaxy model (red dash dotted lines).}
    \label{fig:gc_xzall}
\end{figure*}

Figures \ref{fig:gc_xyall} and \ref{fig:gc_xzall} illustrate the results of our simulations, where we consider the evolution of star clusters and their tidal tails in Galactic potentials with and without dwarf galaxies. Figure \ref{fig:gc_xyall} shows the X-Y positions of tidal tail stars and Figure \ref{fig:gc_xzall} shows the X-Z positions of tidal tail stars in Galactocentric coordinates. For comparison purposes, the progenitor cluster and its orbital path in both the MW and DG galaxy models are illustrated. Based on these simulations, we conclude that the effect that dwarf galaxies have on the evolution of tidal tails can be broken up into three scenarios: 1) differences between tails in the MW and DG models are due to perturbations to the cluster's orbit, 2) differences between tails in the MW and DG models are due to perturbations to the cluster's orbit and individual tail stars, and 3) differences between tails in the MW and DG models are negligible. We will discuss each of these three scenarios, and the cluster's that they apply to, in the following subsections.

\subsection{Perturbations to Cluster Orbits}\label{s_porbit}

In Figures \ref{fig:gc_xyall} and \ref{fig:gc_xzall}, there is an offset between the orbital paths of NGC 4590, Pal 1 and Pal 5 in the MW and DG models. The observed offset can very simply be caused by the mass enclosed within the cluster's orbit changing over time as dwarf galaxies move within the cluster's orbital distance. Since the progenitor clusters have different orbits in the MW and DG models, the tidal tails of each cluster will also be offset from one another. In all three cases, the path of the tails themselves is consistent with the progenitor cluster's orbital path. 

It is interesting note that for Pal 5, the tidal tails in the DG galaxy model also appear to be shorter than the tails in the MW galaxy model. This difference is likely due to the slight change in the cluster's orbital path and phase. More specifically, in the DG galaxy model Pal 5 is closer to apocentre than in the MW galaxy model, which results in the tails being compressed \citep{Kaderali2019}. 

\subsection{Perturbations to Cluster Orbits and Tail Stars}\label{s_pstar}

Similar to the clusters discussed in Section \ref{s_porbit}, the orbital paths of NGC 362, NGC 1851, NGC 4147, NGC 5466, NGC 7492, Pal 14 and Pal 15 in the MW and DG galaxy models are also offset from one another in Figures \ref{fig:gc_xyall} and \ref{fig:gc_xzall}. While differences in the mass enclosed within each cluster's orbit in the DG model will contribute to the orbital path of each cluster being offset from its path in the MW model, a close inspection of the tail stars themselves suggests a secondary mechanism is contributing to the offset.

Taking NGC 5466 as an example, we see in Figure \ref{fig:gc_xyall} that the tail path deviates significantly from the orbital path for stars in the DG model when moving away from the cluster. A smaller deviation is seen in Figure \ref{fig:gc_xyall}. For tail stars with x-coordinates greater than the cluster's x-coordinate, there appears to be a spur-like feature. For tail stars with x-coordinates less than the cluster's x-coordinate, there appears to be a kink in the tail. Both of these features are due to NGC 5466 passing within 24 kpc of he Large Magellanic Cloud at a relative speed of 376.5 km/s approximately 160 Myr ago.


Spurs and kinks can also be seen in NGC 362, NGC 4147, NGC 7492, Pal 14 and Pal 15 to varying degrees. For NGC 1851, stars escape the cluster with a high velocity dispersion, due to the cluster being so massive. Hence the resulting tidal tails are quite spread out, which makes identifying spurs and kinks in the tails very difficult. However it is clear in Figures \ref{fig:gc_xyall} and \ref{fig:gc_xzall} that tail stars in the MW model occupy different regions of the galaxy than tail stars in the DG model while the orbital paths are quite similar in each model, indicating that dwarf galaxies have perturbed the tail stars. Hence direct interactions between tail stars and dwarf galaxies can also contribute to the paths of tidal tails differing between the MG and DG models. This result is consistent with the work of \citet{deBoer20} on interactions between Sagittarius and the GD-1 stream.

\begin{figure}
    \includegraphics[width=0.48 \textwidth]{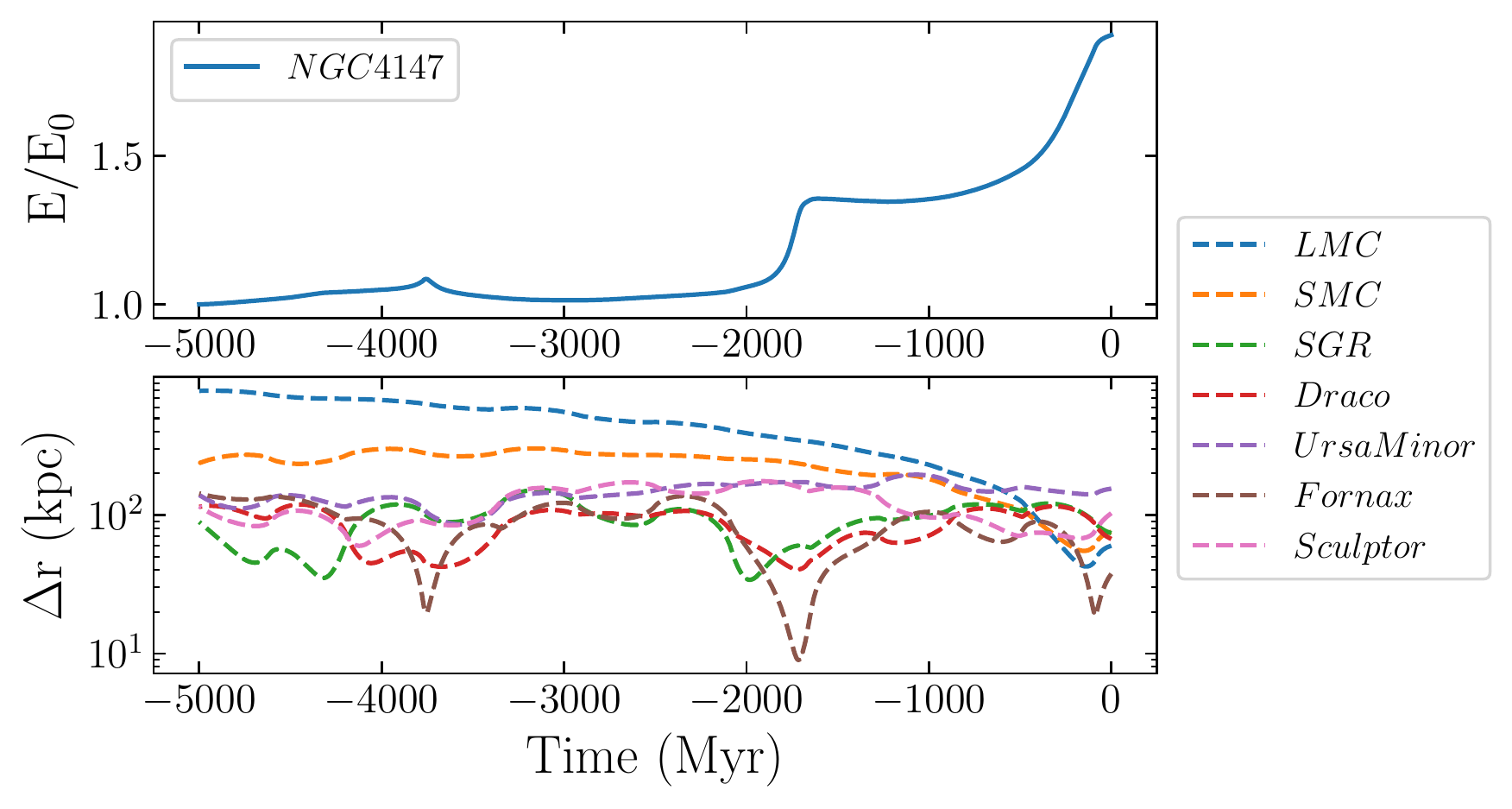}
    \caption{Normalized orbital energy (top panel) and the distances to dwarf galaxies (bottom panel) for NGC 4147. The overall evolution of NGC 4147's orbit energy is driven by the in-fall of the Large and Small Magellanic Clouds. However, interactions with Fornax and Draco also result in abrupt changes to the cluster's orbital energy.}
    \label{fig:ngc4147}
\end{figure}

With the exception of very distinct tail features, like the kink seen in NGC 5466 due to a recent interaction with the Large Magellanic Cloud, it is difficult to attribute specific tail features to individual dwarf galaxy encounters.  For example, Figure \ref{fig:ngc4147} illustrates the evolution of NGC 4147's orbital energy and the relative distance between NGC 4147 and each dwarf galaxy considered in the DG galaxy model. Over the course of 5000 Myr, the orbital energy of NGC 4147 increases due to the in-fall of the Large and Small Magellanic Clouds. However there are several instances, notably 3700 Myr ago and 1700 Myr ago, where interactions with Fornax and Draco result in sharp changes to the cluster's orbital energy. Hence Fornax, Draco, or either of the Magellanic Clouds could be responsible for features in the tails of NGC 4147. If kinks or spurs are observed along a globular cluster's tidal tails, an orbital analysis of individual tail stars is therefore required to determine the dwarf galaxy responsible for creating the feature. For comparison purposes, versions of Figure \ref{fig:ngc4147} for all globular clusters considered in this study are included in the Appendix.

\subsection{Negligibly Affected Tidal Tails}

Finally, the orbital paths and tidal tails of NGC 288, NGC 5139, NGC 5904 and Eridanus show very little offset when simulated in the MW galaxy model and the DG galaxy model. NGC 5139 and NGC 5904 are the two most massive clusters in our dataset and therefore have high escape velocities. Having high stellar escape velocities results in the tails being more spread out, making it difficult to identify specific features in the tails. NGC 5139 is also the innermost clusters in the dataset, with a semi-major axes of 4.3 kpc, which means it will be less affected by dwarf galaxies. Similarly, NGC 288 has a semi-major axis of 7.3 kpc and is therefor not expected to be strongly affected by dwarf galaxies. The only other cluster in our dataset with a semi-major axis within 10 kpc is NGC 362, however as noted in Section \ref{s_pstar} it appears to have had one or more direct encounters with a dwarf galaxy. Figure \ref{fig:a1} in the Appendix confirms NGC 362 has been affected by direct interactions with several dwarf galaxies, including Sagittarius, Fornax, and Draco, in addition to being affected by the current in-fall of the Magellanic Clouds.

The reason for Eridanus being negligibly affected by dwarf galaxies is likely related to a combination of its large semi-major axis (55.8 kpc) and high eccentricity (0.7). The most distant cluster in our dataset is Pal 14, with a semi-major axis of 101.6 kpc and eccentricity of 0.3. Pal 14 has an orbital path that is unaffected by the presence of satellite galaxies. We do however see some spur like features near the edges of the tails due to direct interactions with multiple dwarf galaxies. Figure \ref{fig:a4} in the Appendix illustrates Pal 14 has recently had close passages with Fornax and Sculptor, while also reaching its distance of closest approach to Sagittarius, the Small Magellanic Cloud, and the Large Magellanic Cloud. Hence it appears that the tidal tails of distant clusters are only affected by dwarf galaxies if they have direct interactions, as the change in the enclosed mass within the cluster's orbit is minimal over time. Therefore we conclude that Eridanus has not had any close encounters with a dwarf galaxy over the past 5 Gyr, which explains why its tails are identical in the MW and DG galaxy models.

\subsection{Observing Tidal Tails in Projection}

\begin{figure*}
    \includegraphics[width=\textwidth]{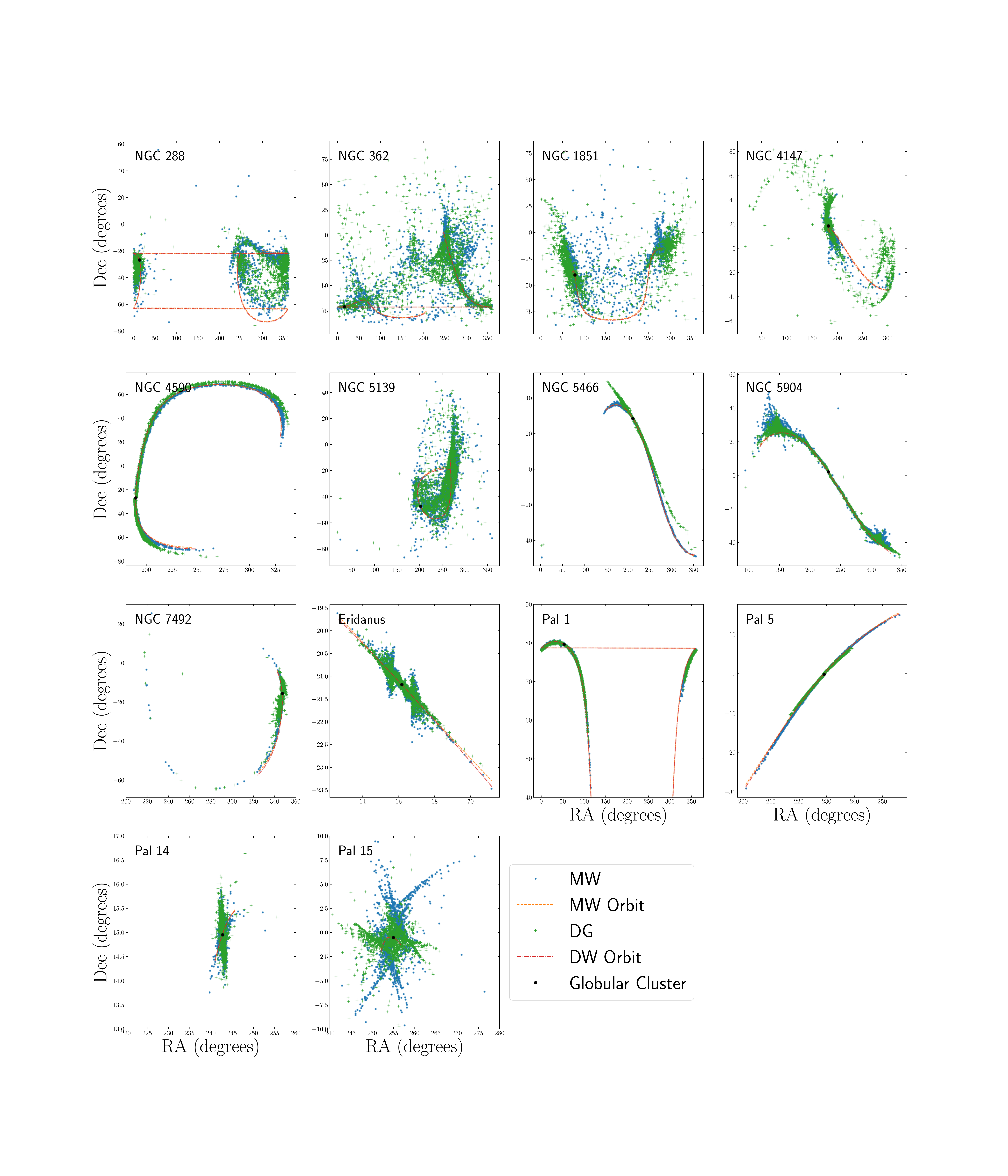}
    \caption{On-sky positions of all simulated tidal tail stars in the MW galaxy model (blue dots) and the DG galaxy model (green crosses) in Galactocentric coordinates. The location of cluster is marked in black, while the orbital path of each cluster is also illustrated for when the cluster orbits in the MW galaxy model (orange dashed lines) and the DG galaxy model (red dash dotted lines).}
    \label{fig:gc_rdall}
\end{figure*}

In order to observationally determine if the tidal tails of a globular cluster have been affected by a dwarf galaxy, measure the properties of dwarf galaxies using tidal tails, or trace the Galactic potential, differences between tidal tails in the MW and DG models must be visible on the plane of the sky or have measured velocity vectors that are significantly misaligned with the stream path \citep{Erkal2019}. We therefore plot the positions of tidal tail stars on the plane of the sky in Figure \ref{fig:gc_rdall}, with the orbital paths of each cluster in the MW and DG models again plotted for comparison purposes.

Figure \ref{fig:gc_rdall} illustrates the previously observed offsets in Galactocentric coordinates between the orbital paths and tidal tails of clusters in the MW and DG galaxy models are more difficult to see on the plane of the sky. For the cases where perturbations to the cluster's orbit lead to an offset between the tidal tails of the cluster in the MW and DG models, offsets are still visible in the plane of the sky for NGC 4590 and Pal 5. However for Pal 5 the offset is small. For Pal 1, the offsets observed in Figures \ref{fig:gc_xyall} and \ref{fig:gc_xzall} are no longer visible on the plane of the sky. Hence accurate distances or measurements of each stars kinematic properties would be necessary to determine that the tails have been perturbed.

For cases where perturbations to the tail stars lead to spurs and kinks in the tidal tails of clusters in the DG model, in addition to differences in the orbital paths of clusters in the MW and DG models, tidal tail offsets are still visible on the plane of the sky for all clusters. However features associated with NGC 7492 and Pal 14 are more difficult to observe without knowledge of each star's kinematic properties.

\section{Discussion}\label{s_discussion}

\subsection{Quantifying the Effect of Dwarf Galaxies on Tidal Tails}

We have demonstrated that including dwarf galaxies in the external potential within which a globular cluster orbits can affect the tidal tails of the cluster. The tail path itself can differ due to the orbital paths of the cluster in the MW and DG galaxy models being different. Alternatively, more direct interactions between tail stars and dwarf galaxies can lead to kinks and spurs. To better quantify how strongly a tidal tail is affected by including dwarf galaxies in the external potential, we consider the dispersion of tail stars about the orbital path of the cluster in the MW galaxy model. The stars that populate the tidal tails of a globular cluster are stars that have escaped the cluster through one of its Lagrange points. Since stars escape the cluster with non-zero velocities, their orbit in the external potential will be slighty different than that of the progenitor cluster. The resulting tidal tail stars will therefore be distributed about the orbital path of the progenitor cluster.

As a measure of how the tidal tails of clusters in the MW galaxy model and DG galaxy model differ, we consider the effect on the distribution of tail stars about the cluster's orbital path in the MW galaxy model. For each tail star in the MW galaxy model, we find the distance to the point along the cluster's orbital path ($D_{path}$) in the MW galaxy model that it is closest to. We then find the standard deviation in $D_{path}$, which we denote as $\sigma_{MW,MW_{path}}$. A similar calculation is done for tail stars in the DG galaxy model with respect to the cluster's orbital path in the MW galaxy model, which we denote $\sigma_{DG,MW_{path}}$. The difference $\sigma_{DG,MW_{path}} - \sigma_{MW,MW_{path}}$ then provides an indication of how much the two tails differ. A difference of zero would indicate the two tidal tail systems are nearly the same, while values not equal to zero indicate that the tidal tail stars in the DW galaxy model are distributed differently about the cluster's tail path in the MW galaxy model. For each set of tidal tails, the difference is calculated using both the three-dimensional and on-sky positions of tail stars with respect to the three-dimensional and on-sky orbital paths of the cluster in the MW galaxy model respectively.

\begin{figure}
    \includegraphics[width=0.48 \textwidth]{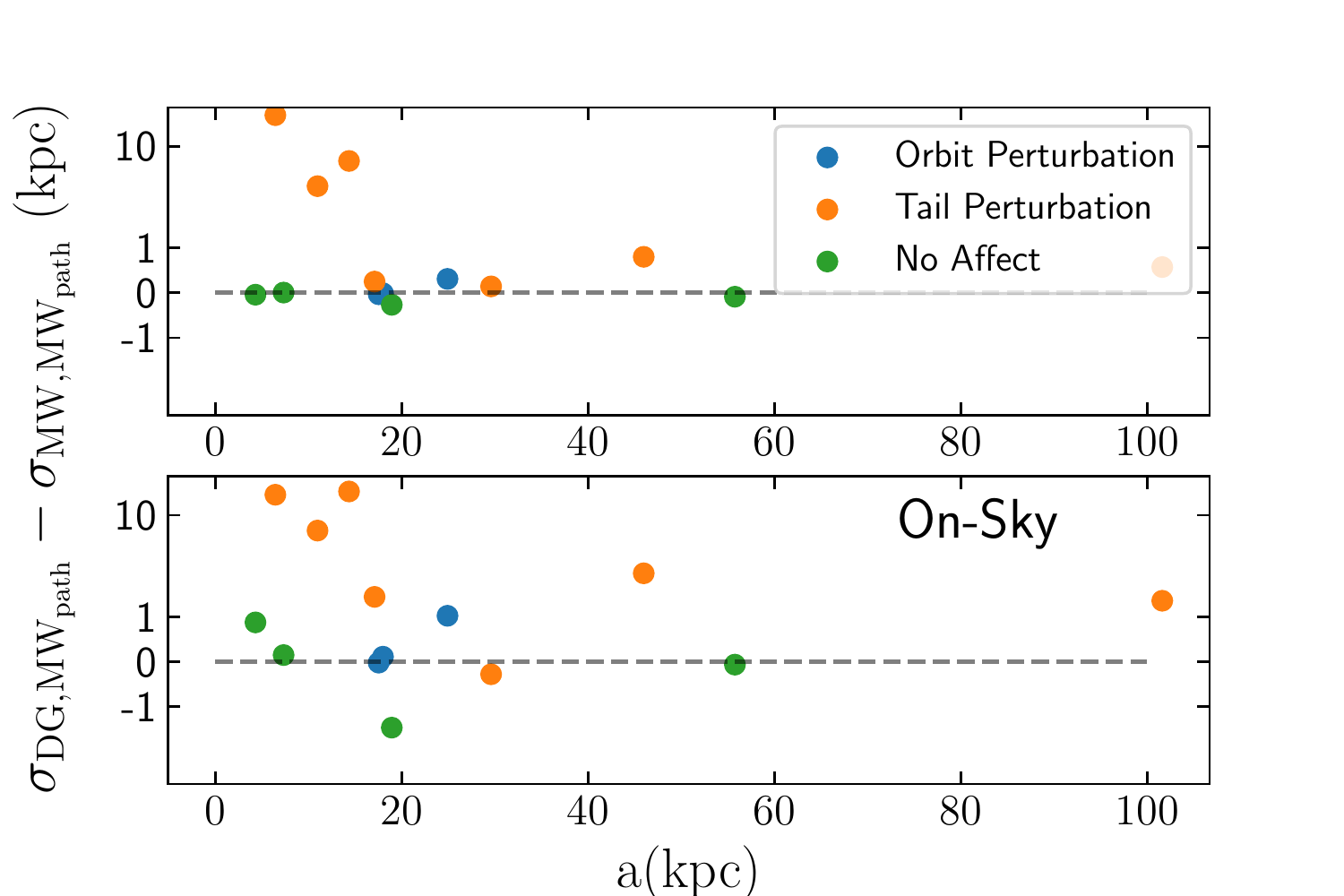}
    \caption{Three-dimensional (top panel) and on-sky (bottom panel) differences between the standard deviation of tidal tail stars in the DG galaxy model about the orbital path of the cluster in the MW galaxy model ($\sigma_{DG,MW_{path}}$) and the standard deviation of tidal tail stars in the MW galaxy model about the orbital path of the cluster in the MW galaxy model ($\sigma_{MW,MW_{path}}$) as a function of each cluster's semi-major axis in the MW galaxy model. Tails are colour coded based on whether or not differences between tails in the MW and DG models are primarily due to perturbations to the cluster's orbit (blue) or direct perturbations to tail stars (orange), or if the tails aren't affected at all by the presence of dwarf galaxies (orange).The dashed horizontal line marks $\sigma_{DG,MW_{path}} - \sigma_{MW,MW_{path}} = 0$. Note the y-axis has a symmetrical log scale.}
    \label{fig:sigma}
\end{figure}

Figure \ref{fig:sigma} illustrates the three-dimensional and on-sky differences $\sigma_{DG,MW_{path}} - \sigma_{MW,MW_{path}}$ for each tidal tail system as a function of the cluster's semi-major axis in the MW galaxy model. The points are colour coded based on our designations in Section \ref{s_results} of whether or not tails appear to differ primarily do to a perturbation to the cluster's orbit in the DG galaxy model, tail perturbations, or if dwarf galaxies have no affect on the tidal tails. Focusing initially on tidal tails in the DG galaxy model that appear to follow the perturbed orbital path of the progenitor cluster, we see $\sigma_{DG,MW_{path}} - \sigma_{MW,MW_{path}}$ is only slightly offset from zero when calculated in three dimensions. NGC 4590 and Pal 5 expectantly show the largest offsets from zero, while $\sigma_{DG,MW_{path}} - \sigma_{MW,MW_{path}}$ for Pal 1 is comparable to the clusters with tails that appear to be unaffected by dwarf galaxies. The result for Pal 1 is understandable as the offset in the orbital paths of the cluster in the MW and DG models is minimal compared to NGC 4590 and Pal 5.

For tails that have had individual stars perturbed by a dwarf galaxy, we see a range in $\sigma_{DG,MW_{path}} - \sigma_{MW,MW_{path}}$ values. Clusters where a large number of stars are part of an observed kink or spur (NGC 362, NGC 1851, NGC 4147) can have $\sigma_{DG,MW_{path}} - \sigma_{MW,MW_{path}} > 3$. Conversely clusters with only minor kink and spur features can still have $\sigma_{DG,MW_{path}} - \sigma_{MW,MW_{path}}$ values that are near zero. Similarly, all four clusters that are visually consistent with not being affected by dwarf galaxies also have  $\sigma_{DG,MW_{path}}-\sigma_{MW,MW_{path}}$ values near zero.

In the lower panel of Figure \ref{fig:sigma}, where $\sigma_{DG,MW_{path}}-\sigma_{MW,MW_{path}}$ is calculated using the on-sky positions of tail stars and the on-sky orbital path of the cluster, we see that projection effects can cause $\sigma_{DG,MW_{path}}-\sigma_{MW,MW_{path}}$ to move closer to or farther from zero depending on the cluster. NGC 4147 appears to be the most strongly affected by projection, as $\sigma_{DG,MW_{path}}-\sigma_{MW,MW_{path}}$ increases from 6.5 to 19.6 when calculated in projection. There is also one case, NGC 5904, where $\sigma_{DG,MW_{path}}-\sigma_{MW,MW_{path}}$ moves away from zero despite appearing to be unaffected by dwarf galaxies when viewed in three dimensions in Figures \ref{fig:gc_xyall} and \ref{fig:gc_xzall}. This apparent change in $\sigma_{DG,MW_{path}}-\sigma_{MW,MW_{path}}$ is artificial and is a result of how the tails and the orbital path are projected onto the plane of the sky. Hence it is necessary to compare observed tails to simulations in order to understand whether projection effects can result in tidal tails appearing to be perturbed by external sources. 

\subsection{Exploring the Effect of the Large Magellanic Cloud }\label{s_Large Magellanic Cloud}

The Large Magellanic Cloud is by far the most massive satellite galaxy of the Milky Way. In fact the Large Magellanic Cloud is nearly four times as massive as the Small Magellanic Cloud, which is the second most massive satellite galaxy. Therefore it is likely that the perturbations to cluster orbits and tail stars observed in our simulations are primarily due to the Large Magellanic Cloud. In fact, in Figures \ref{fig:a1} - \ref{fig:a4} it appears that long-term evolution of each cluster's orbital energy follows the in-fall of the Large and Small Magellanic Clouds. Interactions with other satellites only cause small deviations in each cluster's orbital energy.

\begin{figure*}
    \includegraphics[width=\textwidth]{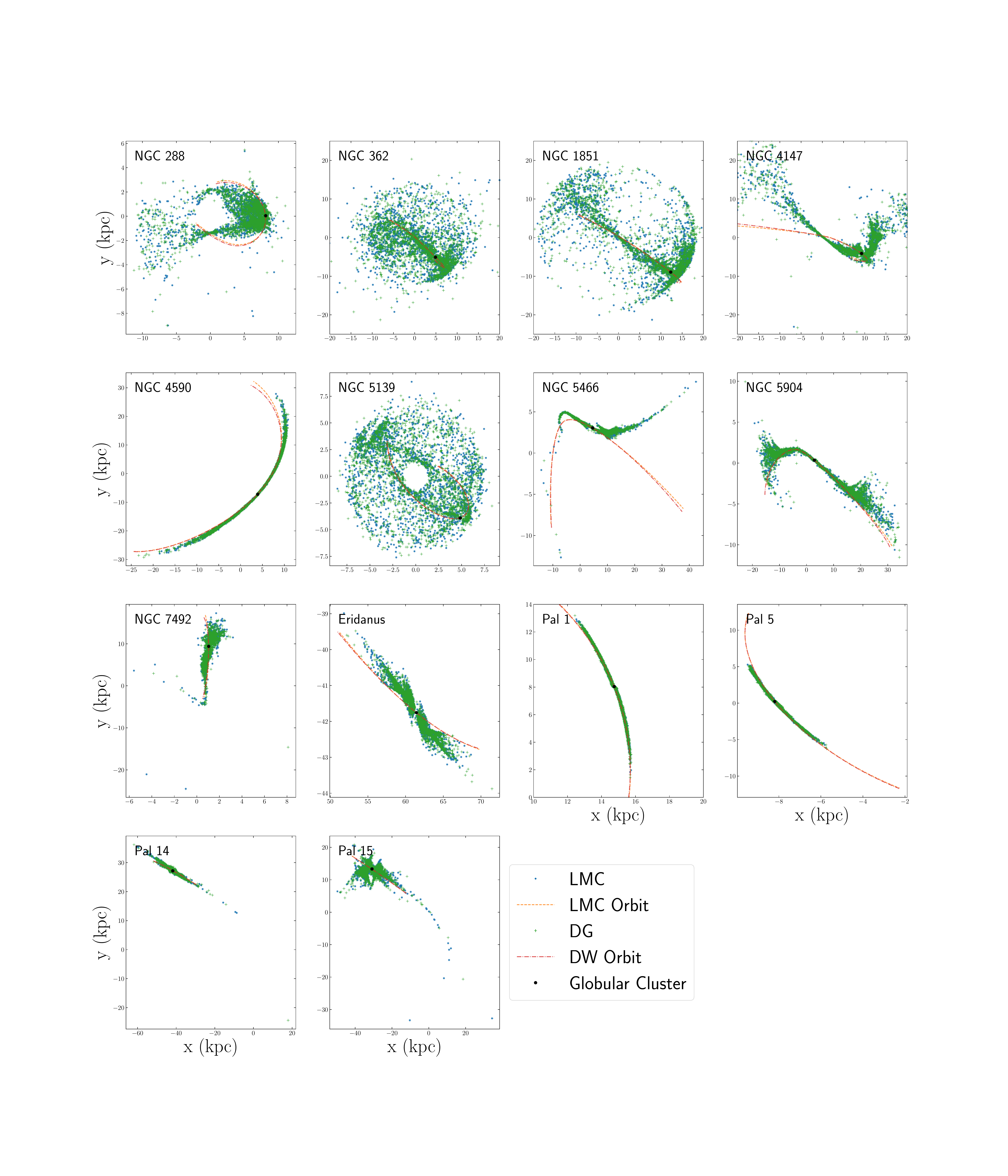}
    \caption{X-Y positions of all simulated tidal tail stars in the LMC galaxy model (blue dots) and the DG galaxy model (green crosses) in Galactocentric coordinates. The location of the cluster is marked in black, while the orbital paths of each cluster is also illustrated for when the cluster orbits in the LMC galaxy model (orange dashed lines) and the DG galaxy model (red dash dotted lines).}
    \label{fig:gc_lmc_xyall}
\end{figure*}

To explore how strongly interactions with the Large Magellanic Cloud dominate tidal tail evolution compared to the entire satellite galaxy population, we construct a third galaxy model (LMC) that only consists of the Milky Way and the Large Magellanic Cloud. Similar to the the DG model discussion in Section \ref{s_methods}, the Milky Way model is set equal to the MWPotential2014 model from \citet{galpy} and the Large Magellanic Cloud is taken to be a Hernquist sphere of mass $1 \times 10^{11} M_{\odot}$ and scale radius 10.2 kpc \citep{Besla10} that has had its orbit integrated in MWPotential2014 while taking into consideration dynamical friction. The results of generating tidal tails for all 14 clusters in the LMC galaxy model with \texttt{streamspraydf} are illustrated in Figure \ref{fig:gc_lmc_xyall} and compared to tails generated in the DG galaxy model. 

Figure \ref{fig:gc_lmc_xyall} reveals that there exists minimal offset between the orbital paths of clusters in the DG and LMC galaxy models. NGC 288, NGC 4147, NGC 4590, NGC 5466, NGC 590, NGC 7492, and Eridanus are the only clusters where the orbital path in the LMC model is remotely offset from the path in the MW model, however in most cases the difference is extremely small.

With respect to the tail stars, there is also no clear difference between the tail paths in either galaxy model. Similarly, the kinks and spurs observed for tidal tails in the DG model also exist for tails in the LMC model. The most significant kink and spur found in this study, as seen in the NGC 5466 tail model, are evident in both the DG and LMC galaxy models. Hence this feature in the tidal tails of NGC 5466 primarily due to one or more close interactions with the LMC that affects tail stars without disrupting the orbital path of the cluster.

\begin{figure}
    \includegraphics[width=0.48 \textwidth]{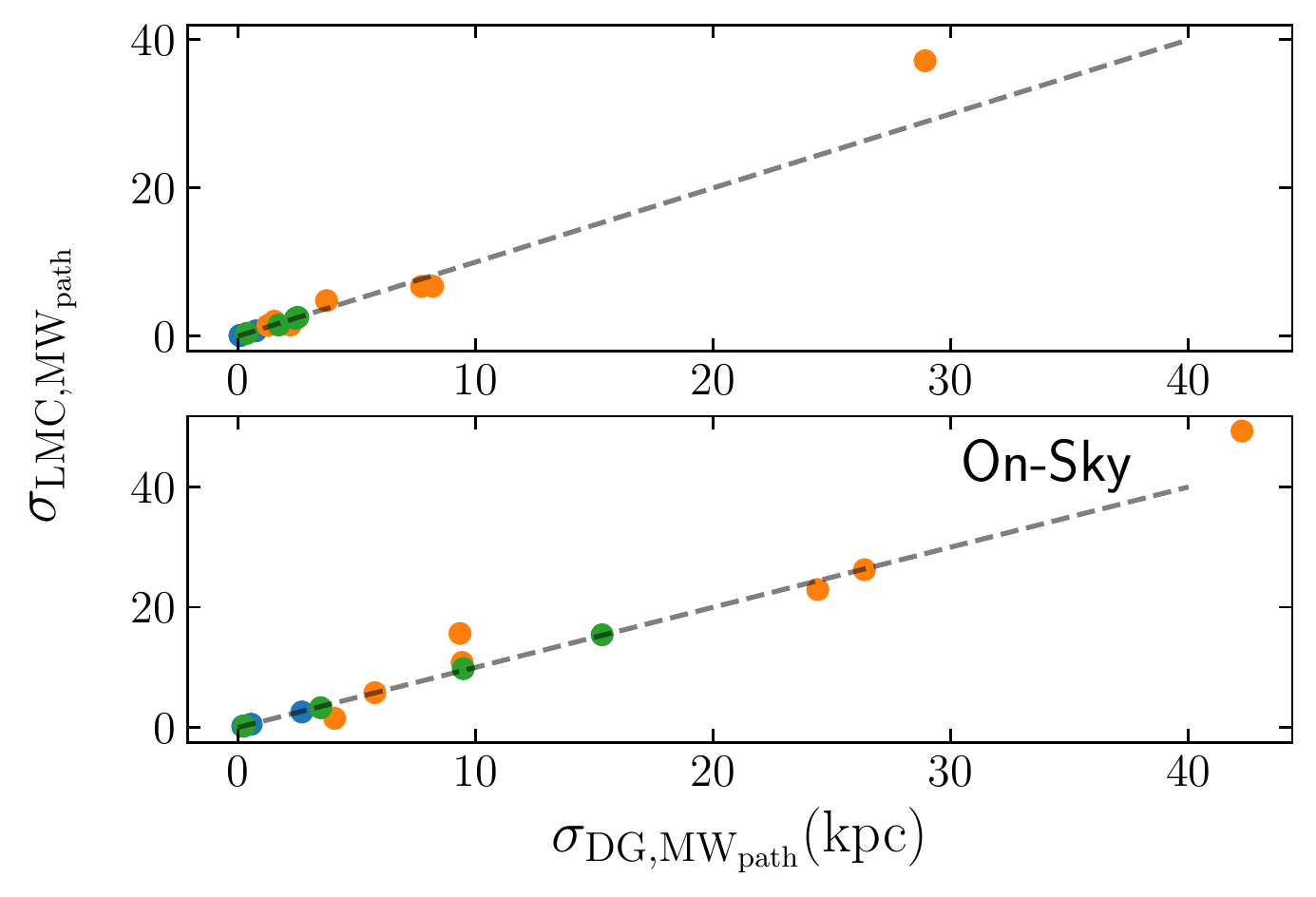}
    \caption{Three-dimensional (top panel) and on-sky (bottom panel) standard deviation of tidal tail stars in the LMC galaxy model about the orbital path of the cluster in the MW galaxy model ($\sigma_{LMC,MW_{path}}$) compared to the standard deviation of tidal tail stars in the LMC galaxy model about the orbital path of the cluster in the MW galaxy model ($\sigma_{DG,MW_{path}}$). Tails are colour coded based on whether or not differences between tails in the MW and DG models are primarily due to perturbations to the cluster's orbit (blue) or direct perturbations to tail stars (orange), or if the tails aren't affected at all by the presence of dwarf galaxies (orange).The dashed horizontal line marks $\sigma_{LMC,MW_{path}} = \sigma_{DG,MW_{path}} = 1$.}
    \label{fig:sigma_lmc}
\end{figure}

To better test the statement that the tidal tails are similar in both the DG and LMC models, we consider our metric for how tidal tail stars are distributed about the cluster's orbital path in the Milky Way. Hence we calculate the the standard deviation in $D_{path}$ for tidal tail stars in the LMC galaxy model using both three-dimensional and on-sky positions, which we denote as $\sigma_{LMC,MW_{path}}$, and compare it to $\sigma_{DG,MW_{path}}$ in Figure \ref{fig:sigma_lmc}.

When comparing $\sigma_{LMC,MW_{path}}$ to $\sigma_{DW,MW_{path}}$ we see that almost all tails fall on the 1:1 line when the standard deviations are calculated using the three-dimensional positions of stars in Galactocentric coordinates. The only exception being NGC 362, where $\sigma_{LMC,MW_{path}}$ is larger than $\sigma_{DG,MW_{path}}$. This difference is likely due to the fact that NGC 362 is a high mass cluster, such that many stars are ejected from the cluster with high velocities and do not remain part of the tails. Non-tail stars that are far from the orbital path can significantly contribute to a change in the standard deviation about the path. Additionally, looking at the orbital energy evolution of NGC 362 in Figure \ref{fig:a1}, NGC 362 is perturbed by several dwarf galaxies in addition to the Large Magellanic Cloud. If we instead calculate the median absolute standard deviation, the values are near identical for NGC 362 tail stars in the DG model and the LMC model.

\section{Conclusion}\label{s_conclusion}

Through the analysis of mock tidal tails generated using a particle spray method, we have explored how dwarf galaxies can affect the tidal tails of the 14 Galactic globular clusters that have observed tidal tails. For each globular cluster, tails are generated in an external field that consists of the Milky Way only (MW galaxy model) and in an external field that consists of the Milky Way and seven of its satellite galaxies (DG galaxy model). When comparing tidal tails generated around the same globular cluster, but in different galaxy models, we find that including dwarf galaxies in the galaxy model can result in the orbital path of the cluster being perturbed, both the orbital path and individual tail stars being perturbed, or there being no affect on the tidal tails. 

The orbital path of a globular cluster in the DG galaxy can deviate from the path in the MW model if a dwarf galaxy passes within the cluster's orbit so the mass enclosed within the cluster's orbit changes. Changing the cluster's orbital path will change the path of cluster's tidal tails, as we observed for the tidal tails of NGC 4590, Pal 1, and Pal 5. A change in the cluster's orbital path can also result in the apparent length of the tails depending on how the cluster's current orbital phase changes as well, as seen for Pal 5. 

Direct interactions between dwarf galaxies and globular cluster can result in a perturbation to the cluster's orbital path and perturbations to individual tail stars. The latter can result in kinks and spurs along the tidal tails, which are seen in the mock tidal tails of NGC 362, NGC 1851, NGC 4147, NGC 5466, NGC 7492, Pal 14 and Pal 15. For NGC 288, NGC 5139, NGC 5904, and Eridanus, the presence of dwarf galaxies in the external tidal field has a negligible affect on tidal tails. A combination of the cluster's orbital distance, orbital eccentricity, and mass is responsible for tails in the MW and DG galaxy models appearing identical for these clusters.

When viewing tidal tails in the MW and DG galaxy models on the plane of the sky, differences sometimes become more difficult to observe. Furthermore, our simulations assume that the cluster's have been disrupting for 5 Gyr. While this assumption was made simply to produce long tidal tails for analysis, it does not necessarily correspond to the observed tails of each system. For cases where an offset between the MW galaxy model tails and the DG galaxy model tails is only visible near the edges of the tails, it may not be observable if the actual tails do not extend to the same distance or if the density of tail stars is too low. Having identified tidal tails that will be affected by dwarf galaxies in this study, detailed simulations must be performed on a cluster-by-cluster basis in order to better compare simulations to observations.


Identifying that the tidal tails of a given cluster will be affected by interactions with dwarf galaxies allows for the properties of satellites galaxies to be constrained by tail observations \citep{Shipp2021}. However, additional factors will need to be considered before directly comparing model tails in different Galaxy models to observations, like the response of the Milky Way itself to its own satellites \citep{GaravitoCamargo19,Garavito2020,Erkal2020, Erkal2021.506.2677E, Petersen2021}, including the Large Magellanic Cloud in mass models of the MW \citep{Erkal2020} and perturbations to globular cluster orbits due to dark matter substructure \citep{Pavanel2021}. Incorporating all of these factors into the evolution of star clusters and their resulting tidal tails will increase our ability to use clusters and tails as tools to study the galaxies within which they reside.

\section*{Acknowledgements}

The authors would like to thank the anonymous referee for some very helpful suggestions on how to improve the manuscript. NE and JW would like to thank the University of Toronto's 2020 Summer Undergraduate Research Program and funding from the Dunlap Institute of Astronomy and Astrophysics, which made this research possible. The authors would also like to acknowledge Jo Bovy and Ting Li for helpful feedback on our project at various stages of its development. NE would like to thank her family and friends for their constant support encouragement. JW would like to thank Denis Erkal for helpful discussions at the Streams 21 Conference hosted by the Flatiron Institute. 

\section*{Data Availability}
The simulated data underlying this article will be shared on reasonable request to the corresponding author.





\bibliographystyle{mnras}
\bibliography{refs} 


\appendix

\newpage

\section{Orbital Energy Evolution of Individual Globular Clusters} \label{s_appendix}

\begin{figure*}
    \includegraphics[width=\textwidth]{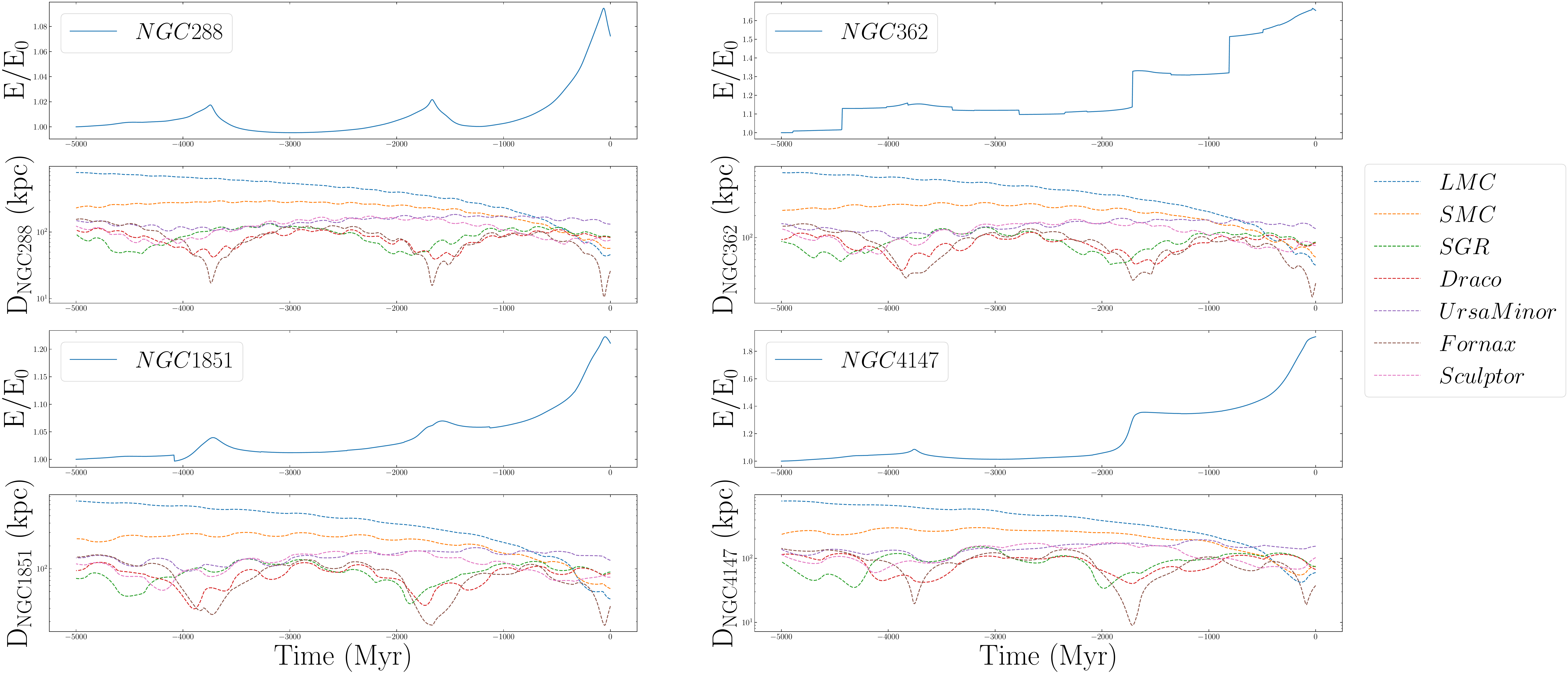}
    \caption{Normalized orbital energies and the distances to dwarf galaxies for globular clusters NGC 288 (top left panels), NGC 362 (top right panels), NGC 1851 (bottom left panels), and NGC 362 (bottom right panels).}
    \label{fig:a1}
\end{figure*}

\begin{figure*}
    \includegraphics[width=\textwidth]{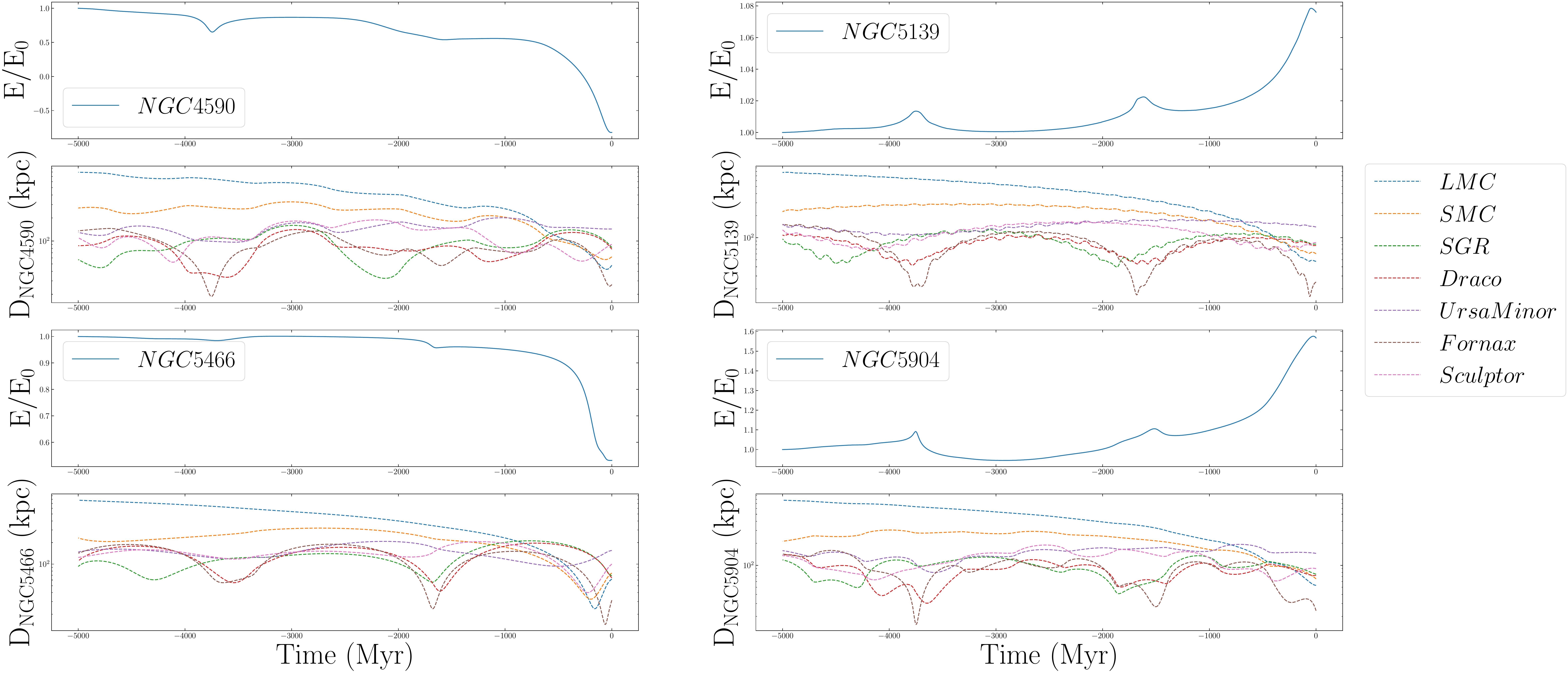}
    \caption{Normalized orbital energies and the distances to dwarf galaxies for globular clusters NGC 4590 (top left panels), NGC 5139 (top right panels), NGC 5466 (bottom left panels), and NGC 5904 (bottom right panels).}
    \label{fig:a2}
\end{figure*}

\begin{figure*}
    \includegraphics[width=\textwidth]{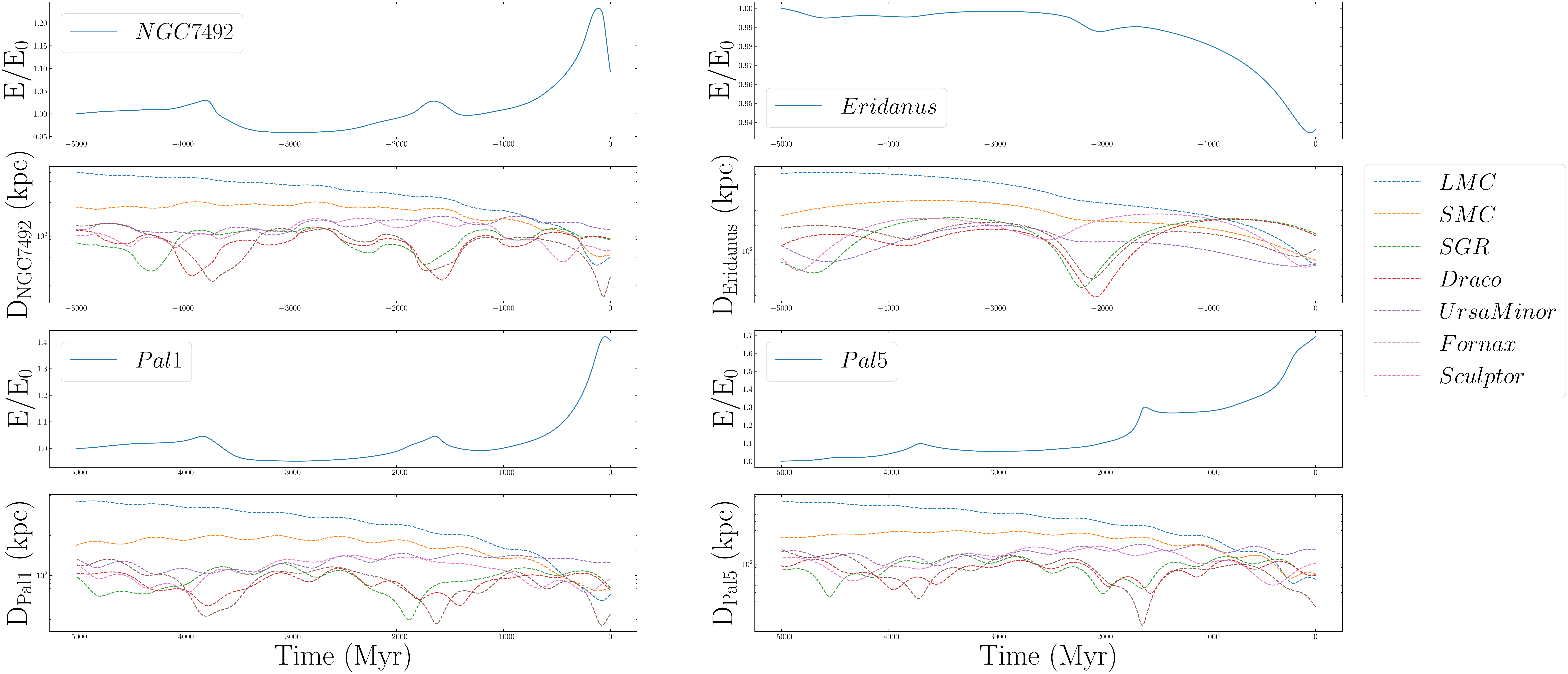}
    \caption{Normalized orbital energies and the distances to dwarf galaxies for globular clusters NGC 7492 (top left panels), Eridanus (top right panels), Pal 1 (bottom left panels), and Pal 5 (bottom right panels).}
    \label{fig:a3}
\end{figure*}

\begin{figure*}
    \includegraphics[width=\textwidth]{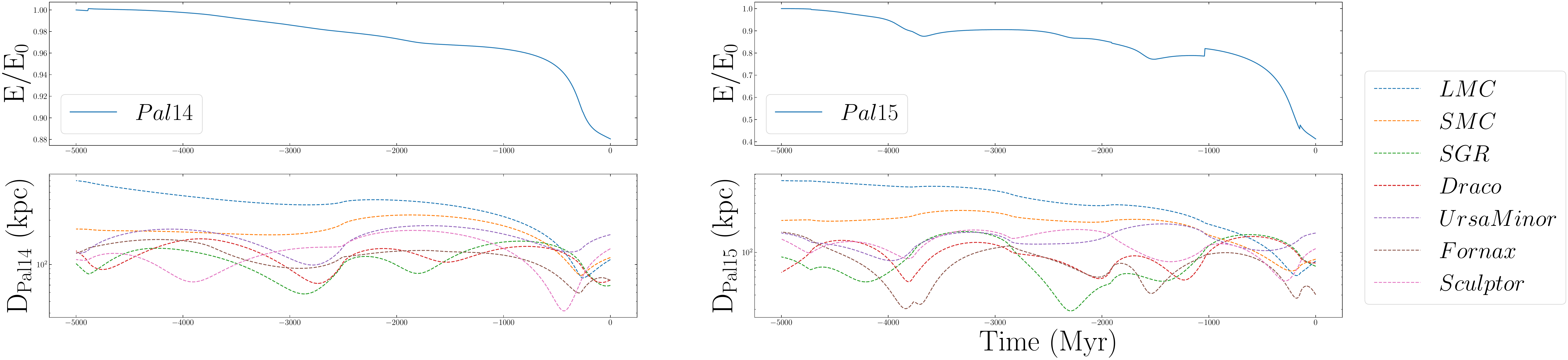}
    \caption{Normalized orbital energies and the distances to dwarf galaxies for globular clusters Pal 14 (left panels) and Pal 15 (right panels).}
    \label{fig:a4}
\end{figure*}

Figures \ref{fig:a1} to \ref{fig:a4} illustrate the orbital energy evolution of each globular cluster considered in this studied. The figures further illustrate the distance between each cluster and every satellite galaxy include in the DG galaxy model. From these figures, it can be seen that while the LMC and SMC dominate the long-term orbital energy evolution of each cluster, interactions with lower mass dwarf galaxies are still capable of altering a cluster's orbit.

\clearpage

\label{lastpage}
\end{document}